\documentclass[10pt,aps,twocolumn,prc,superscriptaddress,showpacs,nofootinbib,noshowkeys,floatfix,preprintnumbers]{revtex4-1}
\usepackage[dvipdfmx]{graphicx}
\usepackage[usenames]{color}
\usepackage{amsmath,amssymb}
\usepackage{multirow}
\usepackage{longtable}
\usepackage[normalem]{ulem}
\usepackage{epstopdf}
\usepackage{times}
\usepackage[normalem]{ulem}  

\renewcommand\sout{\bgroup \color{red} \ULdepth=-.5ex \ULset}
\def\Xint#1{\mathchoice
   {\XXint\displaystyle\textstyle{#1}}%
   {\XXint\textstyle\scriptstyle{#1}}%
   {\XXint\scriptstyle\scriptscriptstyle{#1}}%
   {\XXint\scriptscriptstyle\scriptscriptstyle{#1}}%
   \!\int}
\def\XXint#1#2#3{{\setbox0=\hbox{$#1{#2#3}{\int}$}
     \vcenter{\hbox{$#2#3$}}\kern-.5\wd0}}

\def\dashint{\Xint-}
\usepackage{bm}
\graphicspath{{./Figs/}}
\begin{document}
\title{Sizes of the Nucleon}
\date{\today}
\author{Norbert Kaiser}
\author{Wolfram Weise}
\affiliation{Technical University of Munich,  School of Natural Sciences,  Physics Department, 85747 Garching, Germany}
\affiliation{Excellence Cluster ORIGINS, Boltzmannstr. 2, 85748 Garching, Germany}

\begin{abstract}
Evidences are updated and strengthened for the two-scales picture of low-energy nucleon structure as a compact `hard' valence quark core surrounded by a `soft' cloud of quark-antiquark pairs (the meson cloud).  These considerations are quantified by a spectral analysis of the mean-squared radii associated with the isoscalar and isovector electric form factors of the nucleon.  Further supporting arguments come from corresponding studies of the axial and mass form factors and their inferred radii.  Separating low-mass (mesonic) and high-mass (short-range) contributions in the spectral representations of each of these form factors,  we conclude that a central core with an r.m.s.  radius of about 1/2 fm results consistently as the common feature in all cases.  Implications are discussed for baryonic matter at densities beyond that of equilibrium nuclear matter.
\end{abstract}
\maketitle

\section{Introduction}
Spontaneously broken chiral symmetry in the low-energy limit of QCD governs the long-wavelength structure and dynamics of nucleons.  Pions play a distinguished role in this context as (approximate) chiral Nambu--Goldstone bosons.  The coupling of pions to the nucleon adds a 'soft' surface degree of freedom to its structure.  Models based on chiral symmetry therefore describe the nucleon as a complex system characterized by two scales: a compact 'hard' core and a surrounding quark-antiquark cloud in which pions play a prominent role.  Decades ago chiral quark models of the nucleon (in particular,  the chiral and cloudy bag models \cite{BR1979, Brown1979,MTT1980, TTM1981, Thomas1984}) were designed with this picture in mind~\cite{TW2001}.  Such a delineation between a compact core and a meson cloud \cite{BRW1986} also emerged in descriptions of the nucleon as a chiral soliton (Skyrmion) with vector mesons \cite{Meissner1986, Meissner1987}.  

The idea of a compact core in the center of the nucleon with a size notedly smaller than the proton charge radius was also promoted on the basis of deep-inelastic scattering data at HERA,  together with photoproduction of 
$J/\psi$ and its coherent scattering on nucleons and nuclei \cite{Aktas2006,Caldwell2010}.  Recent evaluations of the nucleon size from high-energy nucleus-nucleus cross sections point in a similar direction \cite{Nijs2022}.

In the meantime the knowledge of nucleon structure in the low-energy,  long-wavelength limit has advanced to a level that does enable a more quantitative evaluation of the core-plus-cloud scenario based on analyses of nucleon form factors and radii.  The present work focuses on three such empirical sources of information: the isoscalar electric charge form factor,  the axial formfactor and the mass formfactor.  The mean-square radii associated with these form factors are all significantly different from each other,  indicating that there is no single 'size' of the nucleon.  However,  by spectral analyses of these form factors,  we collect evidence for a common half-a-fermi sized core inside the nucleon which hosts the three valence quarks and thus the baryon number.  At the same time this core carries most of the nucleon mass generated by the (gluonic) trace anomaly.  The mesonic clouds surrounding this core carry the quantum numbers of the currents which give rise to the respective form factors.  These mesonic surfaces are shown to account for the observed differences in the empirical radii.  

A two-scales structure of the nucleon is supposed to have far-reaching implications for strongly interacting matter at low temperatures and high baryon densities as it is realized in neutron stars \cite{Fukushima2020,Brandes2024}.  This is a prime motivation for investigating empirical constraints on the sizes of central core and mesonic surface regions in the nucleon.

In the present work we argue that the proposed two-scales scenario is indeed manifest in empirical nucleon form factors and corresponding radii. Each form factor $G_\alpha(q^2)$ related to a current operator $J_\alpha^\mu$ with index $\alpha$ (referring e.g.  to the electromagnetic or the axial current) has a representation in terms of an unsubtracted dispersion relation,  
\begin{eqnarray}
G_\alpha(q^2)= \frac{1}{\pi}\int_{t_0}^\infty dt\, \frac{\text{Im} \,G_\alpha(t)}{t-q^2-i\epsilon}~,
\end{eqnarray} 
with the squared four-momentum transfer $q^2=q_0^2 - \vec{q}^{\,2}$.  The normalization $G_\alpha(q^2= 0)$ is identified with the `charge' of the current $J_\alpha^\mu$. 
Mean square radii are then given as
\begin{eqnarray}
\langle r_\alpha^2\rangle = \frac{6}{G_\alpha(0)} \frac{dG_\alpha(q^2)}{dq^2}\Big|_{q^2=0}=
\frac{6}{\pi}\int_{t_0}^\infty\frac{dt}{t^2}S_\alpha(t)~,
\end{eqnarray}
where the distribution $S_\alpha(t)= \text{Im}\,G_\alpha(t)/G_\alpha(0)$ represents the spectrum of intermediate hadronic states through which the external probing field couples to the respective nucleon current.  The low-$t$ regions of these spectral distributions ($t \lesssim t_c \sim 1.1$ GeV$^2$) are expected to be associated with the mesonic surface,  while the high-$t$ range ($t > t_c$) reflects the nucleon core.  We shall now enter into a detailed discussion of three form factors of special interest in this context.

\section{Isoscalar electric form factor and radius}

The nucleon matrix elements $(N=p,n)$ of the electromagnetic current operator,
\begin{eqnarray}
& &\langle N(p')|J^\mu_{em}|N(p)\rangle = \nonumber \\ 
& &\bar{u}(p')\left[F_1(q^2)\gamma^\mu+{i\over 2M}F_2(q^2)\sigma^{\mu\nu}q_\nu\right]u(p)~,
\end{eqnarray}
define the Dirac and Pauli form factors,  $F_1(q^2)$ and $F_2(q^2)$.  The four-momentum transfer is $q^\mu = (p' - p)^\mu$ and $M$ denotes the nucleon mass.  The proton and neutron electric form  factors are given by:
\begin{eqnarray}
 G_E^{p,n}(q^2)= F_1^{p,n}(q^2)+ {q^2\over 4M^2}F_2^{p,n}(q^2)~,
\end{eqnarray}
with charges $G_E^{p}(0) = 1$ and  $G_E^{n}(0) = 0$.  Isoscalar and isovector form factors are given as the combinations:
\begin{eqnarray}
 G_E^{S,V}(q^2)= {1\over 2}\left[G_E^{p}(q^2)\pm G_E^{n}(q^2)\right]~.
\label{eq:GE}
\end{eqnarray}
The slopes of $ G_E^{p,n}$ at zero momentum transfer determine the corresponding mean-squared radii.  The empirical r.m.s.  proton charge radius has been obtained in electron scattering and muonic hydrogen measurements \cite{Pohl2010} reviewed in \cite{Gao2022} and consistently updated in \cite{Lin2022}: $\langle r_p^2\rangle^{1/2} = 0.840\pm 0.003\pm 0.002$ fm. Its combination with six times the slope of the neutron electric form factor,  $\langle r_n^2\rangle = -0.105\pm 0.006$ fm$^2$ \cite{Filin2021},  gives the isoscalar and isovector mean squared charge radii of the nucleon,  $\langle r^2_{S,V}\rangle = \langle r_p^2\rangle \pm \langle r_n^2\rangle$,  resulting in the following values:
\begin{eqnarray}
\sqrt{\langle r_S^2\rangle} = 0.775 \pm 0.011 \,\text{fm}~, \label{eq:radiiS} \\
\sqrt{\langle r_V^2\rangle} = 0.901\pm 0.009 \,\text{fm}~.
\label{eq:radiiV}
\end{eqnarray}
Advanced lattice QCD simulations \cite{Djukanovic2023} have now reached a level of precision that closely approaches these empirical radii.  

The isoscalar electric form factor is a suitable case for discussing a delineation between the `core' and `cloud' parts of the nucleon.  We write it again as an unsubtracted dispersion relation:
\begin{eqnarray}
G_E^S(q^2)=\frac{1}{\pi}\int_{t_0}^\infty dt\, \frac{\text{Im} \,G_E^S(t)}{t-q^2-i\epsilon}~,
\label{eq:DR}
\end{eqnarray} 
normalized as $G_E^S(0) = {1\over 2}$. The spectrum $\text{Im} \,G_E^S(t) = \text{Im} \,F_1^S(t)+{t\over 4M^2}\,\text{Im} \,F_2^S(t)$ with $F_i^S = {1\over 2}(F_i^p + F_i^n)$ starts at the three-pion threshold, $t_0 = 9m_\pi^2$.  It is strongly dominated by the narrow $\omega$ meson while the contribution of the isoscalar $3\pi$ continuum in the range $t\le m_\omega^2$ is negligibly small~\cite{Kaiser2019}.  Additional contributions come from the $\phi$ meson,  its $K\bar{K}$ tail and the $\rho\pi$ continuum,  to be discussed later.

\begin{figure}
\centering
\includegraphics[width=5cm]{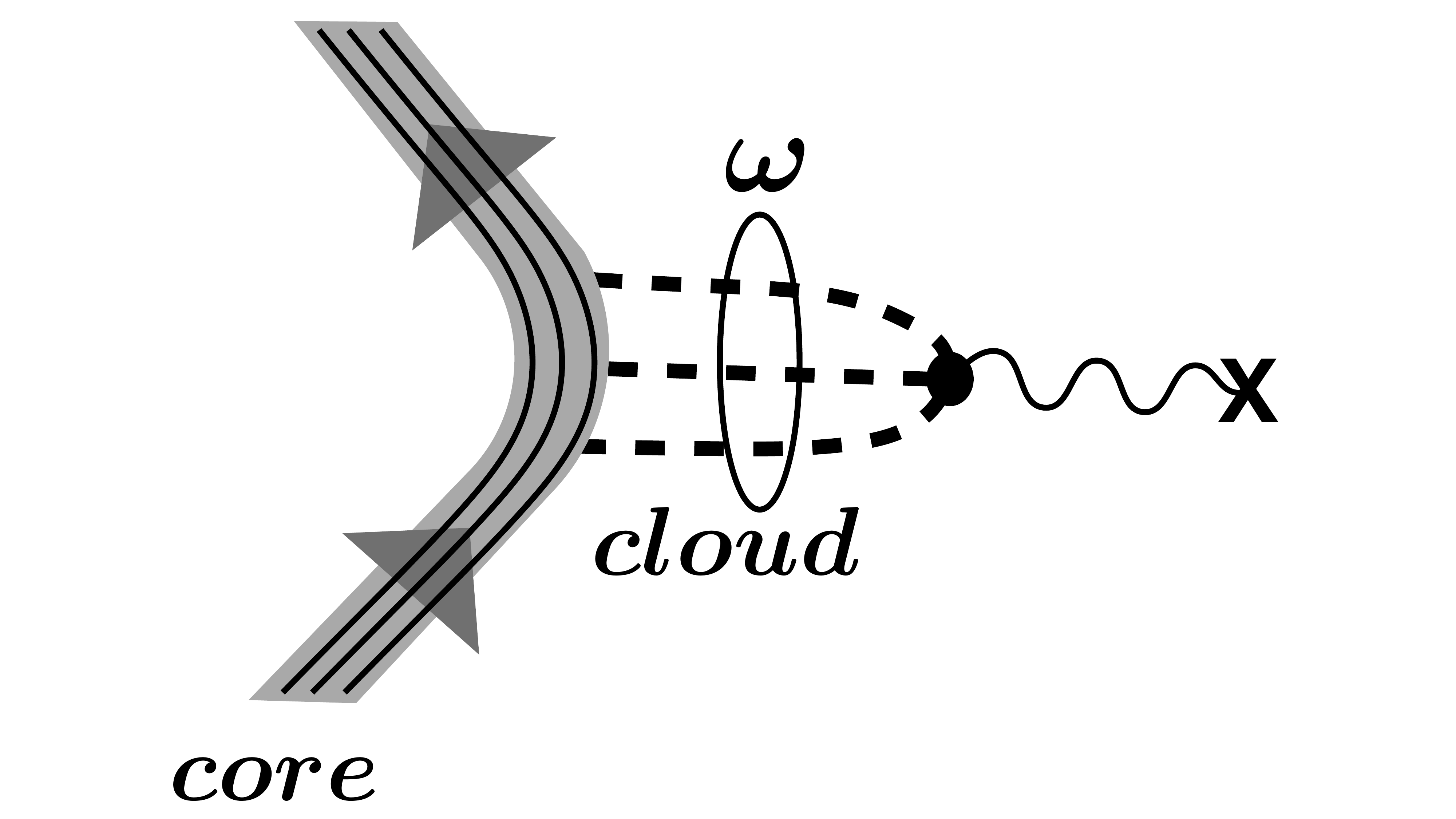}
\caption{Isoscalar electric form factor of the nucleon viewed in a `core plus cloud' picture.  The simplest vector meson dominance model identifies the isoscalar $J^P=1^-$ cloud with the $\omega$ meson.
\label{fig1} }
\end{figure}

For a quick first estimate,  consider the simplest version of a vector meson dominance model (VDM).  In this model the probing isoscalar $J^P=1^-$ photon converts into an omega meson which couples to the nucleon core (see Fig.\,\ref{fig1}).  The flavour SU(3) Gell-Mann - Nishijima formula,  $Q = I_3 + {1\over 2}(B+S)$,  relates the isoscalar charge $Q = {1\over 2}$ to the baryon number $B=1$ for $S=0$.  Therefore the isoscalar charge distribution of the core is also identified with the distribution of baryon number carried by the three valence quarks in the nucleon.  The surrounding quark-antiquark cloud represented by the $\omega$ meson does not contribute to baryon number and electric charge but adds to determining the isoscalar radius, $\sqrt{\langle r_S^2\rangle}$.  In this picture the isoscalar electric form factor is given by the following ansatz:
\begin{eqnarray}
G_E^S(q^2) = \frac{F_B(q^2)}{2(1-q^2/m_\omega^2)}~.
\end{eqnarray}
The form factor $F_B(q^2)$ of the baryon number distribution in the nucleon core (with $F_B(0)=B=1$) acts as the source of the $\omega$ field that propagates with its mass $m_\omega$.  Introducing the mean square radius of the baryon core,  $\langle r_B^2\rangle = 6\frac{dF_B(q^2)}{dq^2}\big{|}_{q^2=0}$,  the mean squared isoscalar charge radius becomes
\begin{eqnarray}
\langle r_S^2\rangle = \langle r_B^2\rangle + \frac{6}{m_\omega^2}~.
\end{eqnarray}
Using $m_\omega = 783$ MeV and the empirical value (\ref{eq:radiiS}) for $\langle r_S^2\rangle^{1/2}$,  the estimated core radius is 
\begin{eqnarray}
\langle r^2_S\rangle^{1/2}_{\text{core}} \equiv \sqrt{\langle r_B^2\rangle} \simeq  0.47\pm 0.01\,\text{fm}~.
\label{eq:VDMradius}
\end{eqnarray}
A nucleon core size of about 1/2 fm is characteristic of chiral `core + cloud' models.  We shall now demonstrate that it  also holds up in a more detailed and realistic treatment \cite{Lin2021,Lin2022} of the spectral distributions governing the isoscalar nucleon electric form factor.  

In what follows we make use of the precision fits to $G_E^S(q^2)$ performed in \cite{Lin2022} for both spacelike and timelike regions of $q^2 = q_0^2 - \vec{q}\,^2$ .  This analysis starts from (\ref{eq:DR})
in such a way that the fitted spectral functions,  $\text{Im}\,F_1^S(t)$ and $\text{Im}\,F_2^S(t)$,  satisfy the normalisation $G_E^S(0)={1\over 2}$ by construction.  The isoscalar mean-squared radius is calculated as:
\begin{eqnarray}
\langle r_S^2\rangle = {12\over\pi} \int_{9m_\pi^2}^\infty dt\left[{\text{Im}\,F_1^S(t)\over t^2} + {\text{Im}\,F_2^S(t)\over 4M^2 t}\right]~.
\label{eq:msradius}
\end{eqnarray} 
\begin{figure}
\centering
\includegraphics[width=7.5cm]{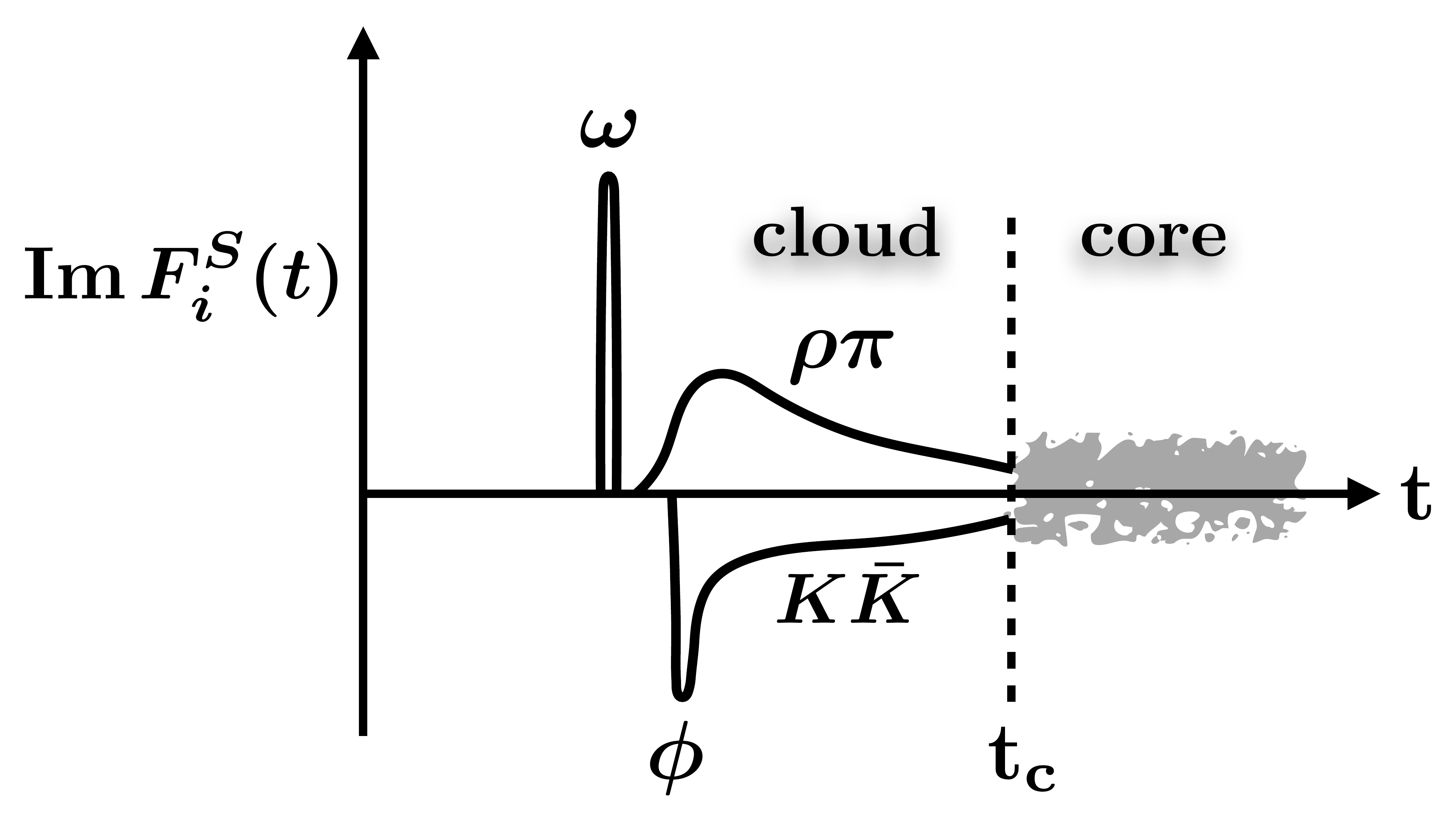}
\caption{Schematic spectral functions of mesonic cloud contributions to the isoscalar electric form factor of the nucleon,  showing $\omega$ and $\phi$ mesons together with $\rho\pi$ and $K\bar{K}$ continuum parts. The higher-mass region $t > t_c$ stands for the nucleon core.  Figure adapted from \cite{Meissner2023}.
\label{fig2} }
\end{figure}
The fits include $\omega$ and $\phi$ meson poles combined with $\rho\pi$ and $K\bar{K}$ continuum contributions as sketched in Fig.\,\ref{fig2}.  The mesonic contributions to the spectral distributions cover a range $t \lesssim t_c \simeq 1.05$\,GeV$^2$,  to be associated with the meson cloud.  The short-distance core part then refers to the region $t > t_c$ and includes information from the timelike domain measured in $e^+e^-\rightarrow N\bar{N}$.  In the analysis of ref. \cite{Lin2022} this region is parametrized by a series of high-mass poles summarized in the Appendix.

The resulting spectral functions are of the form:
\begin{eqnarray}
\text{Im}\,F_{1,2}^S(t)= 
\pi\sum_{i}a_{1,2}^{(i)}\,\delta(m_i^2-t) + \Delta_{1,2}^S(t)_{\text{core}}~,
\label{eq:DR3}
\end{eqnarray} 
where $i=\omega,\phi, ...$ collects the mesonic contributions,  also incorporating the $\rho\pi,K\bar{K}$ continua parametrized effectively in terms of equivalent poles \cite{Belushkin2007}. 
\begin{eqnarray}
 \Delta_{1,2}^S(t)_{\text{core}} = \pi \sum_{j=S_j,\dots}a_{1,2}^{(j)}\,\delta(m_j^2-t)
\end{eqnarray}
refers to the remaining core part in the range $t > t_c$. The parameters of the residua $a_{1,2}^{(i)},  a_{1,2}^{(j)}$ and the pole positions $m_i, m_j$ are listed in Table \ref{tab:Spoles}.
In terms of the latter the isoscalar mean-square core radius is given by:
\begin{eqnarray}
\langle r_S^2\rangle_{\text{core}} = 12\sum_j{a_1^{(j)}\over m_j^4} +{3\over M^2}\sum_i{a_2^{(j)}\over m_j^2} ~.
\label{eq:msradius2}
\end{eqnarray} 
Using the poles $S_j$ and $R_{sj}$ in Table \ref{tab:Spoles},  we find the following result:
\begin{eqnarray}
 \langle r^2_S\rangle_{\text{core}} = (0.237 + 0.017)\,\text{fm}^2 = 0.254 \,\text{fm}^2 \,.
\label{eq:msradius3}
\end{eqnarray} 
The leading number in brackets comes from $a_1^{(j)}$ while the smaller piece refers to $a_2^{(j)}$.  Of the resulting r.m.s. core radius, 
\begin{eqnarray}
 \langle r^2_S\rangle_{\text{core}}^{1/2}\simeq 0.50\, \text{fm} \,,
\label{eq:rms}
\end{eqnarray} 
the by far dominant contribution comes from the poles $S_j$.  The `resonance' poles $R_{sj}$ (which actually include large widths in the original fit of Ref. \cite{Lin2022}),  are of minor importance and contribute less than 2\% to (\ref{eq:rms}).

It is instructive also to take note of the contribution to $\langle r_S^2\rangle$ from the $\omega$ and $\phi$ meson poles: $\langle r_S^2\rangle_\omega + \langle r_S^2\rangle_\phi \simeq 0.589\,\mathrm{fm}^2$.  The remainder of the meson cloud sector comes from the much smaller $\rho\pi$ and $K\bar{K}$ continuum parts.  With inclusion of a (conservative) uncertainty estimate,  
\begin{eqnarray}
\langle r_S^2\rangle^{1/2}_{\text{core}} \simeq 0.50\pm 0.01\,\text{fm}~,
\label{eq:DR3}
\end{eqnarray} 
happens to be remarkably close to the simplest VDM estimate (\ref{eq:VDMradius}).  Hence the 1/2-fermi size scale of the baryonic (valence quark) core in the nucleon,  well distinguished from the much larger charge radius of the proton,  appears to be supported also by an advanced precision fit analysis of the electromagnetic form factors.  

\section{Isovector electric form factor and radius}

The isovector form factor, $G_E^V(q^2)$,  involves the {\it difference} of proton and neutron form factors in (\ref{eq:GE}).  In the limit of perfect isospin symmetry with identical baryonic valence quark cores of proton and neutron,  a first guess would therefore lead to expect that these cores cancel in $G_E^V$ and the isovector core radius should vanish: $\langle r_V^2\rangle_{\text{core}} =0$.

This expectation is confirmed by an inspection of the isovector core radius using the series of fitted poles $V_j, R_{vj}$ in Table\,\ref{tab:Vpoles} \cite{Lin2022},  with the result:
\begin{eqnarray}
 \langle r^2_V\rangle_{\text{core}} =-0.025 \,\text{fm}^2 \,.
\label{eq:msradius4}
\end{eqnarray} 
The small deviation of this value from zero reflects isospin symmetry breaking effects.  The poles $V_j$ are dominant in their magnitudes.  At the same time the different signs of their residua cause the cancellations leading to the almost vanishing balance in (\ref{eq:msradius4}). The `resonance' poles $R_{vj}$ play again only a minor role, contributing 0.003 fm$^2$.

The isovector charge radius of the nucleon thus arises almost entirely from the interacting two-pion cloud \cite{Hoferichter2016} governed by the $\rho$ meson and the prominently enhanced low-mass tail that extends down to the $\pi\pi$ threshold,  $t_0 = 4m_\pi^2$.  The empirical 
\begin{eqnarray}
 \langle r^2_V\rangle =  \langle r^2_V\rangle_{\pi\pi}+ \langle r^2_V\rangle_{\text{core}}= 0.811 \,\text{fm}^2 
\label{eq:msradius5}
\end{eqnarray} 
follows with the $\pi\pi$ continuum and $\rho$ meson cloud contribution,  
\begin{eqnarray}
\langle r^2_V\rangle_{\pi\pi} = 0.836 \,\text{fm}^2~.
\label{eq:msradius6}
\end{eqnarray} 
The observed cancellation of the proton and neutron `core' parts is in essence an indirect confirmation of the two-scales core-plus-cloud structure seen in the analysis of the isoscalar charge radius.

\section{Isovector axial form factor of the nucleon}

As another interesting case we consider next the form factor $G_A(q^2)$ associated with the axial vector current of the nucleon:
\begin{eqnarray}
\langle n|A_-^\mu|p\rangle = \,G_A(q^2)\,\bar{u}_n(p')\, \gamma^\mu\gamma^5\,u_p(p)~.
\label{eq:axialf}
\end{eqnarray} 
It has been deduced \cite{Hill2018} from the weak muon capture process on the proton,  $\mu^-p \rightarrow \nu_\mu n$,  from neutrino scattering on the deuteron and from pion electroproduction,  $e^-p \rightarrow e^-n\pi^+$.  

Given the low-$q^2$ expansion of the axial form factor,
\begin{eqnarray}
G_A(q^2) = G_A(0)\left[1 + {1\over 6}\langle r^2_A\rangle q^2 + \dots\right]~,
\label{eq:axialff}
\end{eqnarray} 
determining the mean-squared radius $\langle r_A^2\rangle$ requires input for the axial vector coupling constant, $g_A = G_A(0)$.  From neutron beta decay,  $g_A = 1.2764(8)$ \cite{Maerkisch2019}.  The extraction from pion electroproduction makes use of the Goldberger-Treiman relation,  $g_A = g_{\pi NN}\,f_\pi/M_n$.  With the pion-nucleon coupling constant $g_{\pi NN} = 13.1$,  the pion decay constant $f_\pi = 92.3$ MeV and the neutron mass $M_n = 939.6$ MeV,  the resulting $g_A$ differs from the empirical value by less than 1\%.  

The matrix elements of the axial current that define $G_A(q^2)$ involve in addition the induced pseudoscalar form factor, $G_P(q^2)$.  It contains the pion pole at $q^2 = m_\pi^2$ and behaves in such a way that PCAC and the GT relation are fulfilled (for details see e.g.  \cite{Hill2018}).

Determinations of $\langle r_A^2\rangle$ reported in \cite{Hill2018} refer to two sources of information: a combined dipole fit to the axial form factor extracted from $\nu d$ scattering and pion electroproduction, which gives $\langle r_A^2\rangle = 0.454\pm 0.013$ fm$^2$,  and a more conservative analysis of $\nu d$ scattering and $\mu p$ capture data, without resorting to an assumed dipole form,  which consequently involves larger uncertainties: $\langle r_A^2\rangle = 0.46\pm 0.16$ fm$^2$.
In either of these two cases,  
\begin{eqnarray}
\langle r_A^2\rangle^{1/2} &=& 0.67\pm 0.01\,\text{fm}\label{eq:RA1}\\
(\text{from}~ \nu d~\text{scattering}~&\text{and}&~e^- p\rightarrow e^-n \pi^+~ \text{dipole fits})~,\nonumber
\end{eqnarray}
\begin{eqnarray}
\langle r_A^2\rangle^{1/2} &=& 0.68\pm 0.11\,\text{fm}\label{eq:RA2}\\
(\text{from}~ \nu d~ \text{scattering} ~&\text{and}&~\mu p~\text{capture}\, \text{analysis})~,  \nonumber
\end{eqnarray}
the axial radius is evidently smaller than the proton charge radius by about 20 \%.

\begin{figure}
\centering
\includegraphics[width=5cm]{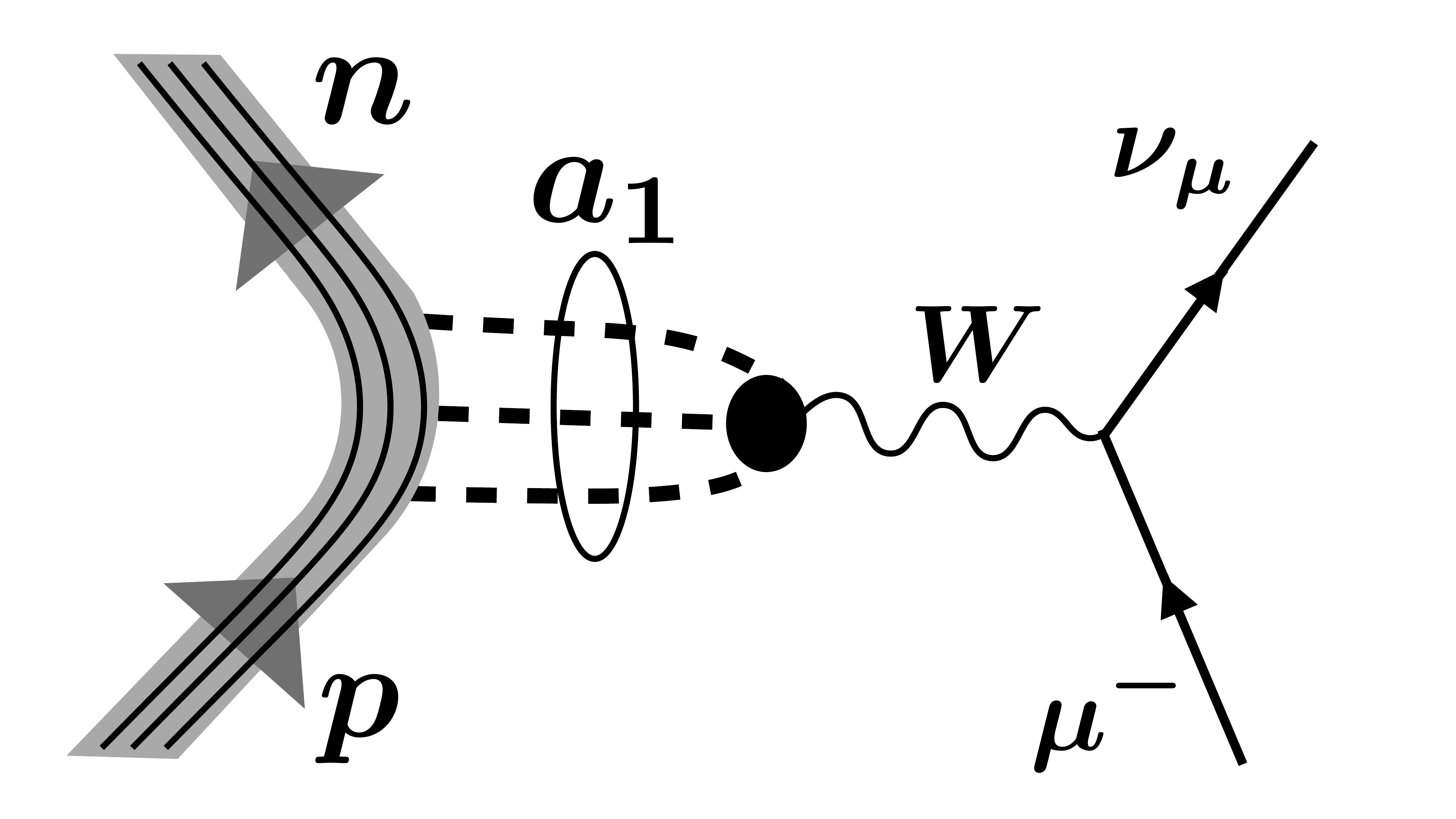}
\caption{Axial form factor of the nucleon as measured in muon capture on the proton and sketched in a 'core plus cloud' picture.  The meson cloud is dominated by the isovector $J^P=1^+$ three-pion spectrum in which the $a_1$ meson figures prominently.
\label{fig3} }
\end{figure}

Writing the axial formfactor as an unsubtracted dispersion relation,
\begin{eqnarray}
G_A(q^2)= \frac{1}{\pi}\int_{t_0}^\infty dt\, \frac{\text{Im}\,G_A(t)}{t-q^2-i\epsilon}~,
\label{eq:axialDR}
\end{eqnarray}
and recalling the normalisation $G_A(q^2=0)= g_A$,  the corresponding mean-squared radius is:
\begin{eqnarray}
\langle r_A^2\rangle &=& \frac{6}{g_A} \frac{dG_A(q^2)}{dq^2}\Big|_{q^2=0}\nonumber\\
&=&\frac{6}{g_A \pi}\int_{t_0}^\infty\frac{dt}{t^2}{\text{Im}\,G_A(t)}~.
\label{eq:Raxial}
\end{eqnarray}
The isovector $J^P=1^+$ spectrum,  Im\,$G_A(t)$,  starts at the three-pion threshold, $t_0=9m_\pi^2$,  and prominently features the broad $a_1$ meson resonance as sketched in Fig. \ref{fig3}.

Let us start again with a simple estimate using a schematic axial vector dominance picture.  It assigns the leading part of the surface contribution to $G_A(q^2)$ through the spectrum of the $a_1$ meson (with its large width).  An approximate scale of this `cloud' part can be introduced by an $a_1$ pole,  Im\,$G_A(t) = g_A\pi\delta(t-m_a^2)$, with a mass $m_a\simeq 1.2$ GeV.  Using the empirical $\langle r_A^2\rangle$ one finds for the remaining `core' size:
\begin{eqnarray}
\langle r_A^2\rangle_\text{core}^{1/2} = \left(\langle r_A^2\rangle - {6\over m_a^2}\right)^{1/2} 
\simeq 0.54\pm 0.01\,\text{fm} ,
\end{eqnarray}
if the dipole fit value (\ref{eq:RA1}) is taken for reference.  Using (\ref{eq:RA2}) instead the uncertainty in $\langle r_A^2\rangle_\text{core}^{1/2}$ increases to about 25\%.  

\begin{figure}
\centering
\includegraphics[width=10cm]{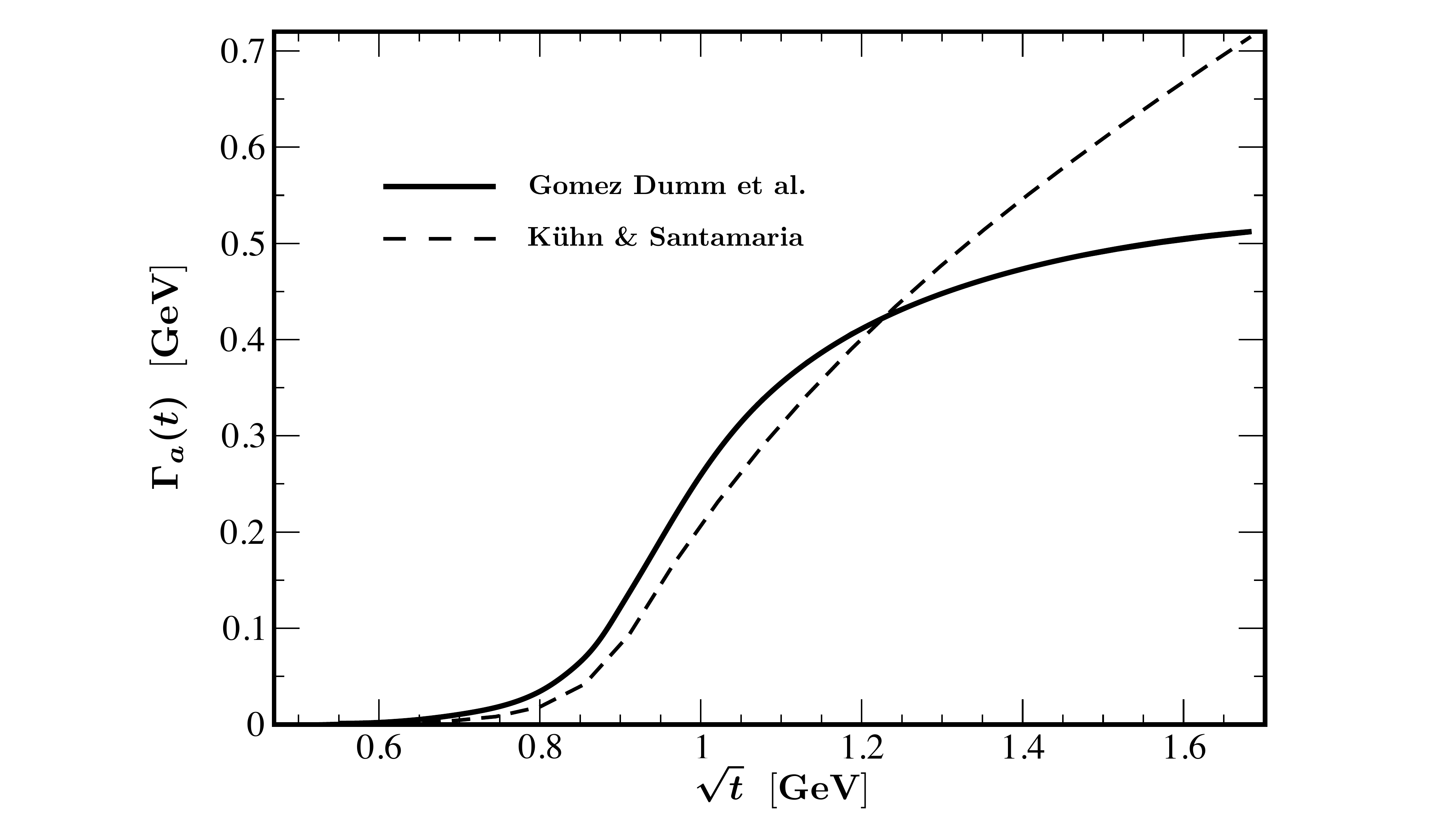}
\caption{Energy dependence of the $a_1$ width $\Gamma_a(t)$ according to \cite{Kuehn1990} (dashed line) and \cite{Gomez2010} (solid line).  The input $a_1$ mass is $m_a=1.23$ GeV and the on-shell width is $\Gamma_a(m_a^2) = 0.425$ GeV (PDG mean values \cite{PDG2022}).  
\label{fig4} }
\end{figure}

A more detailed evaluation requires full account of the broad isovector $J^P=1^+$ three-pion spectral distribution.  We start from the ansatz:  
\begin{eqnarray}
G_A(q^2) = {g_A\,m_a^2\over m_a^2-q^2+ \Sigma_a(q^2) -\text{i}\,m_a\,\Gamma_a(q^2)}~.
\label{eq:GA}
\end{eqnarray}
The real self-energy correction $\Sigma_a(q^2)$,  compatible with the dispersion relation (\ref{eq:axialDR}),  is determined by a twice-subtracted dispersion relation:
\begin{eqnarray}
\Sigma_a(q^2) = {q^2\over\pi}(q^2-m_a^2)~\dashint_{9m_\pi^2}^\infty {dt\over t}{m_a\,\Gamma_a(t)\over (t-m_a^2)(t-q^2)}~,\nonumber\\
\label{eq:selfE}
\end{eqnarray}
where the subtractions leave $g_A$ and $m_a$ untouched.
Results from $\tau\rightarrow \pi\pi\pi\nu_\tau$ decays can be used to set constraints on the energy dependence of the $a_1$ width,  $\Gamma_a(t)$.  In the present work we employ the widths shown in Fig.\,\ref{fig4} taken from \cite{Kuehn1990,Gomez2010}.  In the latter work \cite{Gomez2010} the $a_1\rightarrow\rho\pi\rightarrow 3\pi$ amplitude is integrated over the three-pion phase space,  the information needed in order to identify the meson-cloud sector of $G_A(q^2)$.  With this input the principal value integral in (\ref{eq:selfE}) is performed.  This is done over a limited range, $9m_\pi^2\leq t \leq t_u$,  with the upper limit chosen symmetrically as $t_u = 2m_a^2-9m_\pi^2$ for simple practical reasons.  Taking the derivative of (\ref{eq:GA}) at $q^2 = 0$,  the $a_1$ contribution to the squared axial radius is given by: 
\begin{eqnarray}
\langle r_A^2\rangle_{a_1} = {6\over m_a^2}\left(1+\delta_a\right)~,
\end{eqnarray}
with the correction term
\begin{eqnarray}
\delta_a = -{m_a^3\over\pi}\,\dashint_{9m_\pi^2}^{t_u}dt{\Gamma_a(t)\over t^2(t-m_a^2)}
\end{eqnarray}
depending on the chosen energy-dependent $a_1$ width.  

As an example,  setting $m_a = 1.23$ GeV and using the energy-dependent width from \cite{Gomez2010} shown by the full line in Fig.\,\ref{fig4},  one finds $\delta_a = 0.12$ and $\langle r_A^2\rangle_{a_1} = 0.173\,\text{fm}^2$,  so that 
\begin{eqnarray}
\langle r_A^2\rangle_\text{core}^{1/2} \simeq 0.53\pm 0.02\,\text{fm}~, 
\label{eq:RAcore}
\end{eqnarray}
with an estimated uncertainty based on (\ref{eq:RA1}) and a correspondingly larger one if (\ref{eq:RA2}) is used. The alternative choice \cite{Kuehn1990} of the $a_1$ width gives $\delta_a = 0.04$ and a slightly larger core radius of $0.54$ fm,  still consistent within the uncertainties. Variations of $m_a$ by $\pm\,3\%$ and of $\Gamma_a$ by about  $\pm\,40\%$ as indicated by the PDG values for the $a_1(1260)$ \cite{PDG2022} lead to only marginal changes well within the uncertainties in (\ref{eq:RAcore}).

The core radius (\ref{eq:RAcore}) deduced from the axial formfactor is less accurately determined than the core radius (\ref{eq:DR3}) extracted from the analysis of the isoscalar electric form factor.  It is nonetheless remarkable that,  starting from two independent form factors with quite different empirical r.m.s.  radii,  the separation of the mesonic parts of the respective spectral functions from the high-$t$ sections consistently reveals a common half-fermi scale for the core size inside the nucleon.  

\section{Mass form factor and radius}

\begin{figure}
\centering
\includegraphics[width=5cm]{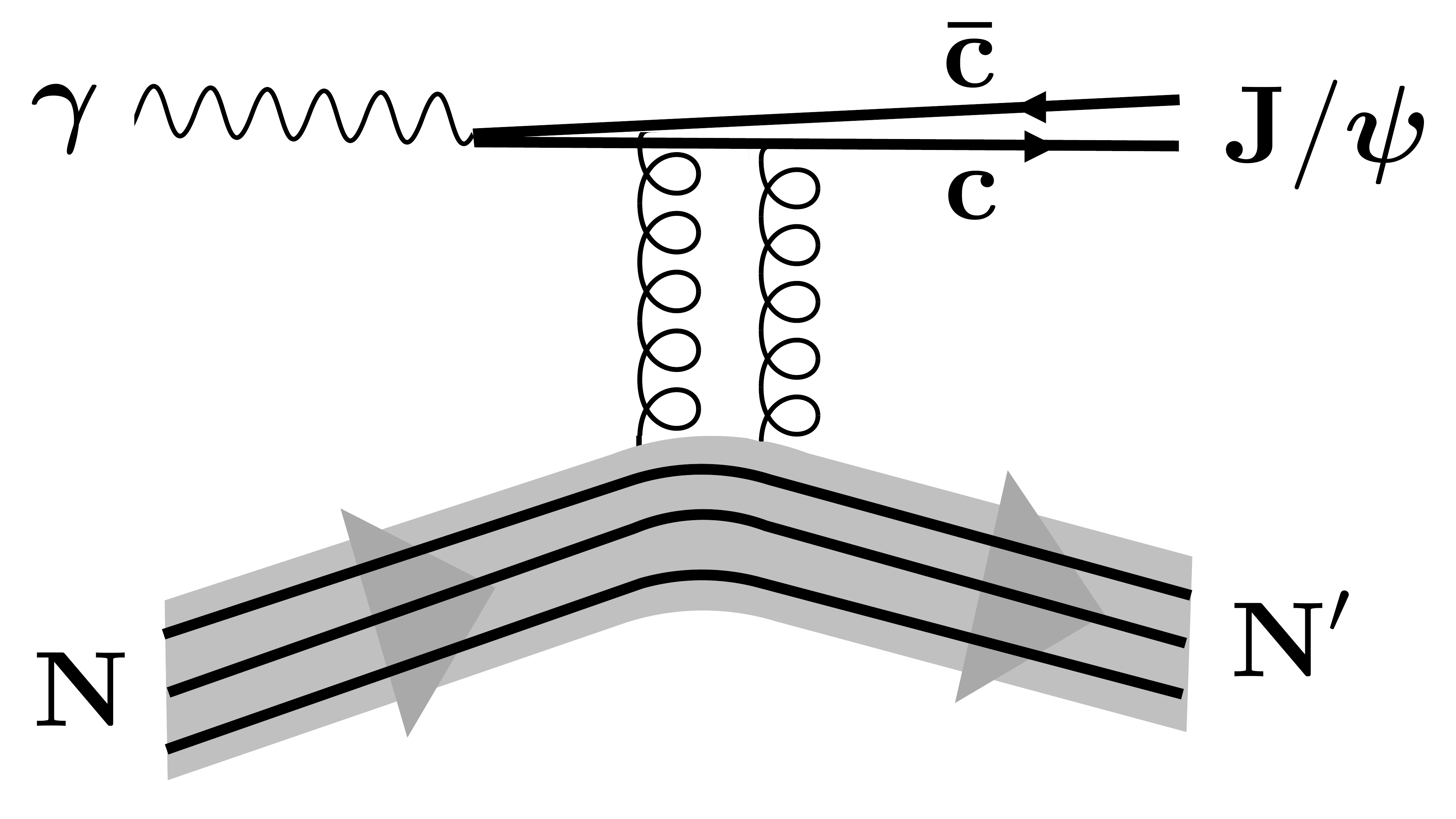}
\caption{Photoproduction of $J/\psi$ with leading two-gluon exchange as a probe of the gluonic structure of the nucleon and its mass form factor.
\label{fig5} }
\end{figure}

\begin{figure}
\centering
\includegraphics[width=8cm]{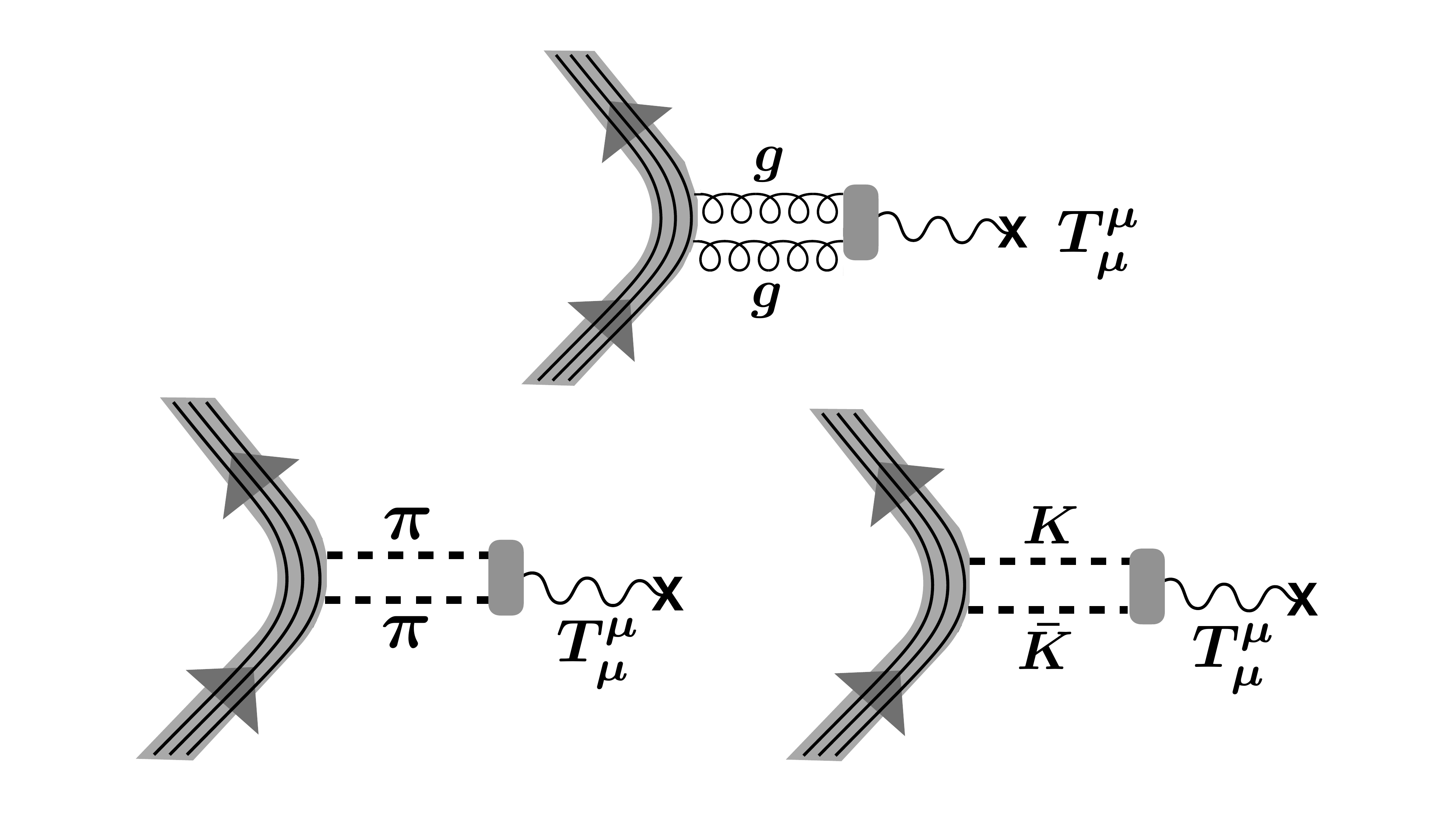}
\caption{Contributions to the spectrum,  $\text{Im}\,G_m(t)$,  of the nucleon's mass form factor: leading two-gluon exchange component (upper diagram),  $\pi\pi$  and $K\bar{K}$ contributions (lower diagrams).
\label{fig6} }
\end{figure}

A further interesting quantity in this context is the mass radius of the proton deduced from $J/\psi$ photoproduction data~\cite{Caldwell2010, Kharzeev2021}.  The $c\bar{c}$ pair forming the $J/\psi$ acts as a small dipole (see Fig.\ref{fig5}) that couples to the nucleon through leading two-gluon exchange in QCD \cite{Kharzeev1999,Martin2000, Frankfurt2001}.  As shown in \cite{Kharzeev2021} the amplitude for this process close to the $J/\psi$ production threshold is proportional to the matrix element of the trace,  $T_\mu^\mu$,  of the nucleon's energy-momentum tensor: 
\begin{eqnarray}
{\cal M}_{\gamma\,N\rightarrow \psi\,N'} \simeq  -{32\pi^2 eCM\over 3b}\langle N(p')|T_\mu^\mu|N(p)\rangle~,
\label{eq:JPsi}
\end{eqnarray}
where $C$ is the coefficient describing the coupling of the gluon fields to the small-sized $c\bar{c}$ pair, $M$ is the nucleon mass and $b = 11 - 2N_f/3 = 9$ for $N_f = 3$ light quark flavors.
This result is valid in the chiral limit of massless quarks, with $T_\mu^\mu = -{bg^2\over 32\pi^2}G_{\mu\nu}^a G^{\mu\nu a}$.  The complete expression includes small additional terms involving the light quark masses\footnote{Heavy ($c, b$ and $t$) quarks appear only as virtual $Q\bar{Q}$ loops in gluon propagators.  Their mass terms in $T_\mu^\mu$ cancel against corresponding heavy-quark sectors in the gluon term.}  and defines the mass (or `gravitational') form factor,  $G_m(q^2)$,  of the nucleon, with $q^2 = (p' - p)^2$:
\begin{eqnarray}
& &G_m(q^2) = \langle N(p')|T_\mu^\mu|N(p)\rangle \nonumber\\
&=& \langle N(p')|{\beta(g)\over 2g}G_{\mu\nu}^a G^{\mu\nu a} + m_l(\bar{u}u +\bar{d}d)+m_s\bar{s}s|N(p)\rangle~.
\nonumber \\
\label{eq:Gmass}
\end{eqnarray}
Here $\beta(g) = -{bg^3\over 16\pi^2}$ is the leading QCD beta function,  $m_l = {1\over 2}(m_u + m_d)$ is the average of the light $u$- and $d$-quark masses,  and $m_s$ is the mass of the strange quark\footnote{These quark masses are actually understood to include the corresponding anomalous mass dimensions.}.  The three terms in (\ref{eq:Gmass}), 
\begin{eqnarray}
G_m(q^2) = G_m^{(0)}(q^2) + \sigma_{N}(q^2) + \sigma_s(q^2)~,
\end{eqnarray}
are identified with the gluonic form factor,
\begin{eqnarray}
G_m^{(0)}(q^2) = \langle N(p')|{\beta(g)\over 2g}G_{\mu\nu}^a G^{\mu\nu a}|N(p)\rangle~,
\end{eqnarray}
and the scalar form factors,
\begin{eqnarray}
\sigma_{N}(q^2) &=& \langle N(p')|m_l(\bar{u}u +\bar{d}d)|N(p)\rangle~, \\
\sigma_s(q^2) &=& \langle N(p')|m_s\bar{s}s|N(p)\rangle~.
\label{eq:sigmaff}
\end{eqnarray}
They represent the pieces illustrated in Fig.\,\ref{fig6} from gluon-dominated short-distance structures,  $\pi\pi$ and $K\bar{K}$ continuum contributions, respectively.

A once-subtracted dispersion relation representation of the mass form factor,
\begin{eqnarray}
G_m(q^2) = M +{q^2\over\pi} \int_{t_0}^\infty{\text{Im}\,G_m(t)\over t(t-q^2 - i\epsilon)}~,
\label{eq:Gm}
\end{eqnarray}
displays the normalisation to the nucleon mass,  $G_m(0) = M$.  By far the largest part of $M$ is generated by gluon dynamics through the trace anomaly (the gluonic term in $T_\mu^\mu$).  The quark mass contributions are given by the pion-nucleon and strangeness sigma terms,  
\begin{eqnarray}
\sigma_{N}\equiv  \sigma_{N}(q^2=0)\quad\text{and}\quad\sigma_s\equiv \sigma_s(q^2=0)~.
\label{eq:sigma}
\end{eqnarray}
In the overall sum, 
\begin{eqnarray}
M = M_0 + \sigma_{N} + \sigma_s~,
\end{eqnarray}
$M_0$ refers to the `core' mass generated by the gluonic trace anomaly,  while the sigma terms account together for less than 10\% of the total nucleon mass $M$. 

The $J^P = 0^+$ two-gluon system couples strongly to the scalar-isoscalar two-pion continuum.  The lower limit in the spectral integral (\ref{eq:Gm}) is therefore at $t_0 = 4m_\pi^2$.  Unlike the prominent low-mass $\pi\pi$ spectrum with $J^P = 1^-$ in the isovector electric form factor,  the $\pi\pi$ continuum contribution to $G_m(q^2)$ is however suppressed in this case because of the small ratio $\sigma_{N}/M$.  

The squared radius of the mass distribution,
\begin{eqnarray}
\langle r_m^2\rangle = \frac{6}{M} \frac{dG_m(q^2)}{dq^2}\Big|_{q^2=0}=
\frac{6}{M \pi}\int_{t_0}^\infty\frac{dt}{t^2}{\text{Im}\,G_m(t)}~,\nonumber \\
\label{eq:Rmass}
\end{eqnarray}
has been extracted from the differential $J/\psi$ photoproduction cross section measured by GlueX \cite{Ali2019, Adhikari2023, Winney2023}.  The result quoted in~\cite{Kharzeev2021},
\begin{eqnarray}
\langle r^2_m \rangle^{1/2} =  0.55\pm 0.03\,\text{fm}~,
\label{eq:Rm}
\end{eqnarray}
is based on a dipole fit to $d\sigma(\gamma p\rightarrow \psi p)/dq^2$ in the form $G_m(q^2) = M(1-q^2/\Lambda^2)^{-2}$.  The resulting $\Lambda = 1.24\pm 0.07$ GeV translates into the radius $\langle r_m^2\rangle^{1/2}=\sqrt{12}/\Lambda$.

˝The assumed proportionality to $G_m(q^2)$ of the coherent $J/\psi$ production amplitude (\ref{eq:JPsi}) relies on  leading $t$-channel two-gluon exchange at small $q^2$.  Alternative mechanisms have been discussed in the literature \cite{Winney2023},  such as open-charm coupled-channel  loops involving $D,  D^*$ and $\Lambda_c$ intermediate states \cite{Du2020}.  In the $t$-channel exchange processes relevant to the slope of $d\sigma/dt$ at leading $q^2$, such heavy-mass objects would enter at scales corresponding to distances far below the $\langle r_m^2\rangle^{1/2}$ given in Eq.\,(\ref{eq:Rm}). They are therefore counted as processes taking place deeply inside the `core' region,  even if they may contribute significantly to the $\gamma + p \rightarrow J/\psi + p$ total cross section.

The result (\ref{eq:Rm}) for $\langle r^2_m \rangle^{1/2}$ is inferred from data restricted to the threshold region of  $J/\psi$ production.  At the same time it is notable that an analysis of diffractive $J/\psi$ electroproduction at high energies,  systematically including both coherent and incoherent processes \cite{Mantysaari2016,Traini2019},  points to characteristic proton size scales with values comparable to (\ref{eq:Rm}).

The elements of the spectral distribution illustrated in Fig. \ref{fig5},  namely short-distance two-gluon exchange plus longer range $\pi\pi$ and $K\bar{K}$ components,  imply the following decomposition of the mean-squared mass radius:
\begin{eqnarray}
\langle r^2_m \rangle= {M_0\over M}\langle r^2_0 \rangle+{\sigma_{N}\over M}\langle r^2_{\pi\pi} \rangle+{\sigma_s\over M}\langle r^2_{K\bar{K}} \rangle~.
\label{eq:Rm2}
\end{eqnarray}
\\
The dominant gluonic trace anomaly contribution with mass $M_0 \simeq 860$ MeV is identified with the squared `core' radius,  $\langle r^2_0 \rangle \equiv \langle r^2_m \rangle_{\text{core}}$, while the small corrections from $\pi\pi$ and $K\bar{K}$ `cloud' pieces involve the sigma terms (\ref{eq:sigma}).  With the aim of estimating the core size, $\langle r^2_m \rangle_{\text{core}}$,  based on the empirical value (\ref{eq:Rm}) of the mass radius,  we proceed now with a discussion of the sigma terms and the associated radii in (\ref{eq:Rm2}).

For the sigma term $\sigma_{N}$ there is as yet not a fully consistent picture.  Recent Lattice QCD computations \cite{Agadjanov2023} give $\sigma_{N} = 43.7\pm 1.2\pm 3.4$ MeV.  A similar result was found in \cite{Liu2021}.   Such values are close to the result obtained decades ago in the time-honored work of \cite{Gasser1991}: $\sigma_{N} = 45\pm 8$ MeV,  together with a large radius of the isoscalar s-wave $\pi\pi$ distribution in the nucleon surface,  $\langle r^2_{\pi\pi} \rangle^{1/2} \simeq 1.3$ fm.  For a derivation of the  scalar-isoscalar form factor of the nucleon see also \cite{Hoferichter2012}.  More recent evaluations \cite{Hoferichter2015, Hoferichter2023} based on a detailed analysis of updated pion-nucleon scattering data have raised this sigma term to $\sigma_{N} = 55.9\pm 3.5$ MeV.  The origin of this larger value of $\sigma_{N}$ is interpreted in \cite{Hoferichter2023} as being related to virtual excited resonance states in $\pi N$ scattering.  A similar value,  $\sigma_{N} = 57\pm 7$ MeV was extracted from a large-scale fit to pionic atom data \cite{Friedman2019}. Both observations are at variance with the LQCD result and the previous determination.  In the following evaluation we can take these differences as a rough measure of possible uncertainties.  

The strangeness sigma term $\sigma_s$ is taken from LQCD  \cite{Agadjanov2023}: $\sigma_s = 28.6\pm 6.2\pm 3.5$ MeV.  An estimate of the corresponding radius of the $K\bar{K}$ cloud can be plausibly obtained by assuming that the size scales of the mesonic $\pi\pi$ and $K\bar{K}$ surfaces are related to the inverse masses of the corresponding thresholds in the spectral function: $\langle r^2_{K\bar{K}} \rangle = (m_\pi/m_K)^2 \langle r^2_{\pi\pi} \rangle$.  In any case the strange quark contribution to the mass radius turns out to play only a very minor role.

With this input an estimate of the `core' radius can be performed.  Using in (\ref{eq:Rm2}) the values $\sigma_{N} \simeq 50$ MeV,  $\langle r^2_{\pi\pi} \rangle \simeq 1.6$ fm$^2$ and the quantities in the strangeness sector as indicated,   the radius of the compact gluonic core of the nucleon that contains most of its mass becomes:
\begin{eqnarray}
\langle r^2_m \rangle_{\text{core}}^{1/2}\equiv\langle r^2_0 \rangle^{1/2} =  0.48\pm 0.05\,\text{fm}~,
\label{eq:Rcore}
\end{eqnarray}
once again close to the common `1/2-fermi rule' for the nucleon core regions observed in the isoscalar electric and axial form factors analysed in this study.  The LQCD values for $\sigma_N$ and $\sigma_s$ including their errors,  as well as possible variations of $\sigma_N$ between 40 and 60 MeV,
are covered by the uncertainty range given in (\ref{eq:Rcore}).

\section{Summary and Conclusions}

Analyses of spectral functions in dispersion relations have been performed to extract radii from the electric, axial and mass form factors of the nucleon. The aim is to delineate the size scales associated with the quark-gluon core from those of a quark-antiquark surface.  Evidence is found supporting a picture of the nucleon as containing a compact half-fermi sized `hard'  core in which the three valence quarks with their baryon number are confined.  This core also hosts the dominant part of the nucleon mass,  the one driven by gluons and the trace anomaly.  It is surrounded by a `soft' surface of quark--antiquark pairs forming the mesonic clouds.  The spectral distributions of these mesonic clouds are characterized by the quantum numbers of the underlying nucleon currents.  They account for the variety of r.m.s.  radii associated with the different respective form factors.  

It should be pointed out that while the localized distributions of charge,  axial current and mass in the nucleon are frame-dependent,  the mean-squared radii discussed in the present work are well-defined frame-independent quantities as they are given by the slopes of invariant form factors at $q^2 = 0$.  They can therefore serve as characteristic size scales independent of a given frame of reference.

The results are summarized as follows:

(i) For the isoscalar electric form factor of the nucleon the separation of core (high-$t$) and mesonic (low-$t$) sectors in the spectral function yields an r.m.s. core radius
\begin{eqnarray}
\langle r^2_S \rangle_{\text{core}}^{1/2} =  0.50\pm 0.01\,\text{fm}~.\nonumber
\end{eqnarray}
The $q\bar{q}$ cloud parts, dominated by the $\omega$ and $\phi$ mesons and supplemented by $\rho\pi$ and $K\bar{K}$ continuum contributions, establish a squared meson-cloud radius $\langle r^2_S \rangle_{\text{cloud}} \simeq 0.35\,\text{fm}^2$ such that the empirical isoscalar charge radius,  $\langle r^2_S \rangle^{1/2} =\sqrt{\langle r^2_S \rangle_{\text{core}}+\langle r^2_S \rangle_{\text{cloud}}}\simeq 0.78$ fm,  is reproduced.  
These results are based on a precise parametrisation of form factors measured in both space- and timelike domains.  The core radius deduced from the isoscalar electric form factor (which coincides with the radius of the baryon number distribution) is in fact the most accurately determined one of all core radii analysed in the present study.  An additional successful test is provided by the isovector electric form factor in which the individual proton and neutron core parts are expected to cancel in the limit of exact isospin symmetry.  

(ii) The radius empirically extracted from the form factor associated with the isovector axial current of the nucleon
is significantly smaller than the proton charge radius.  However,  after separating the broad three-pion spectrum dominated by the $a_1$ meson from the dispersion relation representation of this form factor,  the remaining `core'   part reveals once again a radius compatible with an approximate half-fermi scale: 
\begin{eqnarray}
\langle r^2_A \rangle_{\text{core}}^{1/2} \simeq  0.53\, \text{fm}~,\nonumber
\end{eqnarray} 
with an uncertainty of about $\pm 4\%$ if the empirical dipole fit of  $\langle r^2_A \rangle$ is taken for reference (and a correspondingly larger error if the dipole constraint is released). 

(iii) A third independent source of information is the squared mass radius, $\langle r_m^2\rangle$, derived from the nucleon matrix element of the trace of the QCD energy-momentum tensor.  This information is accessible in the forward differential cross section for near-threshold photoproduction of the $J/\psi$.  It is dominated by the leading short-range two-gluon exchange mechanism between the color-dipole $c\bar{c}$ pair and the nucleon,  and it receives additional contributions from long-range scalar-isoscalar $\pi\pi$ and $K\bar{K}$ components.  The latter corrections are weighted by the pion-nucleon and strangeness sigma terms,  $\sigma_N$ and $\sigma_s$,  which measure the small $u, d$ and strange quark contributions to the nucleon mass $M$.  By far the largest part comes from the gluonic trace anomaly which generates more than 90\% of $M$.  Subtracting estimates of the $\pi\pi$ and $K\bar{K}$ cloud contributions from the empirical $\langle r_m^2\rangle$ one arrives at a radius of the central core in the nucleon which hosts almost all of its mass:
\begin{eqnarray}
\langle r^2_m \rangle_{\text{core}}^{1/2} \simeq  0.48 \pm 0.05\, \text{fm}~.\nonumber
\end{eqnarray} 
The uncertainty includes the error from a dipole fit to the $\gamma+p\rightarrow J/\psi+p$ differential cross section together with possible variations of the sigma terms.

In summary,  the striking feature of all three investigated form factors is the approximate equality of the extracted nucleon core radii:
\begin{eqnarray}
\langle r^2_S \rangle_{\text{core}}^{1/2} \simeq \langle r^2_A \rangle_{\text{core}}^{1/2} \simeq \langle r^2_m \rangle_{\text{core}}^{1/2} \equiv R_{\text{core}}\simeq  {1\over 2}\, \text{fm}~.
\label{eq:Rcore}
\end{eqnarray} 
By its shared properties with the different underlying currents and operators,  this core encloses at the same time the baryon number (i.e.  the three valence quarks) and almost all of the nucleon mass (i.e.  its gluonic trace anomaly part).  In particular the combined spectral analyses of the isoscalar and isovector electric form factors imply that the baryon number $B=1$,  identified with twice the isoscalar charge,  is entirely located in the compact core.  Together with the size information from the gluon-dominated mass form factor this suggests indeed that the three valence quarks,  dressed by strong gluon fields,  are confined within the half-fermi core.

The soft quark-antiquark clouds which form the nucleon surface differ in their mesonic quantum numbers and thus account for the differences e.g.  in observed charge,  axial and mass radii.  In the case of the mass radius,  the large size of the scalar-isoscalar two-pion cloud is down-scaled by being weighted with the small nucleon sigma term which gives only a few-percent correction to the total nucleon mass,  and this therefore explains why the empirical $ \langle r^2_m \rangle$ is close to its gluonic `core' part.

With a common core size $R_\text{core} \sim 1/2$ fm of Eq. (\ref{eq:Rcore}) and a cloud range typically around $R_\text{cloud} \sim 1$ fm,  there is a significant separation of volume scales for a nucleon in vacuum: $(R_\text{cloud}/R_\text{core})^3 \sim 8$.  This two-scales scenario has implications for the behavior of nucleons in dense baryonic matter.  The hard-core and soft-surface sectors of the nucleons behave differently with increasing baryon density.  
At $\rho \simeq \rho_0 = 0.16\, \text{fm}^{-3}$,  the density of equilibrium nuclear matter,  the tails of the meson clouds of nucleon pairs overlap,  resulting in two-body exchange forces.  As the average distance between nucleons decreases with increasing density the soft clouds of $q\bar{q}$ pairs expand and mediate many-body forces involving larger numbers of nucleons.  Their strength increases with characteristic powers of density.  The compact cores,  on the other hand,  are expected to stay intact and isolated until they begin to touch and overlap at average nucleon-nucleon distances $d \lesssim 1$ fm,  corresponding to baryon densities $\rho \gtrsim 6\,\rho_0$.  Note also that random close packing  \cite{Zaccone2022} of hard spheres with a radius $R = 0.5$ fm takes place at a density $\rho\simeq 8\,\rho_0$.  The overlapping of nucleon cores and the deconfinement of valence quarks proceeds at a high cost of energy: further compression of baryonic matter still has to overcome the strongly repulsive short-distance hard core in the nucleon-nucleon interaction. 

As an outlook,  recent analyses of neutron star data \cite{Brandes2023,  Koehn2024} including the heaviest so far observed pulsar (the 2.3 solar-mass black widow pulsar PSR J0952-0607),  require a sufficiently stiff neutron star matter equation-of-state.  As a consequence,  baryon densities reached in the cores of even very heavy neutron stars do not exceed about five times the density of normal nuclear matter.  With the suggested scale separation between a compact 1/2 fm valence quark core and a surface mesonic cloud,  the valence quarks from overlapping cores would be  released,  if at all,  only in the deep interior of extremely heavy neutron stars.  

\subsection{Appendix}
\label{sec:App}

Here we summarize the positions and residua of the poles used in the precision fits of Ref.\,\cite{Lin2022} to the isoscalar and isovector combinations of electric proton and neutron form factors,  both in the spacelike and timelike regions of squared four-momentum transfer $q^2$. Tables \ref{tab:Spoles} and \ref{tab:Vpoles} collect the high-t poles that represent the short-distance `core' sectors in the nucleon.  These parameters determine the core radii given in Eqs.\,(\ref{eq:DR3},\ref{eq:msradius4}).  Table \ref{tab:Spoles} also includes the parameters of the $\omega$ and $\phi$ meson poles used in the evaluation of the isoscalar meson cloud.  

The parameters denoted $S_i$ and $V_i$ refer to zero-width poles.  In the original fits of  Ref.\,\cite{Lin2022} the high-mass resonance poles $R_{si}$ and $R_{vi}$ also have large widths but their effects on the radii analysed in the present work are marginal so that these widths can be ignored in practice.

\begin{table}
\renewcommand{\arraystretch}{1.5}
\setlength{\tabcolsep}{10pt}
\begin{center}
\caption{Parameters for the meson and core sectors of the spectral distribution (\ref{eq:DR3}) representing the isoscalar nucleon form factor $G_E^S(q^2)$.  (Adapted from \cite{Lin2022}).}
\label{tab:Spoles}
\begin{tabular}{|c||c|c|c|}  
\hline 
 $i$ & $a_1^{(i)} ~[\mathrm{GeV}^2]$ & $a_2^{(i)} ~[\mathrm{GeV}^2]$ & $m_i~[\mathrm{GeV}]$\\ \hline
$\omega$ & $0.701$ & $0.338$ & $0.783$\\
$\phi$ & $-0.526$ & $-0.997$ & $1.019$\\ \hline
 $j$ & $a_1^{(j)} ~[\mathrm{GeV}^2]$ & $a_2^{(j)} ~[\mathrm{GeV}^2]$ & $m_j~[\mathrm{GeV}]$\\ \hline
$ S_1$ & $0.422$ & $3.655$ & $1.120$\\
$S_2$ & $0.122$ & $-0.228$ & $1.019$\\
$S_3$ & $0.955$ & $-1.122$ & $1.827$\\ \hline
$R_{s1}$ & $4.953$ & $0.501$ & $1.879$\\
$R_{s2}$ & $-2.653$ & $-1.753$ & $1.903$\\
$R_{s3}$ & $-3.069$ & $2.017$ & $1.914$\\
\hline
\end{tabular}
\end{center}
\end{table}

\begin{table}
\renewcommand{\arraystretch}{1.5}
\setlength{\tabcolsep}{10pt}
\begin{center}
\caption{Parameters for the core sector of the spectral distribution (\ref{eq:DR3}) representing the isovector nucleon form factor $G_E^V(q^2)$.  (Adapted from \cite{Lin2022}). }
\label{tab:Vpoles}
\begin{tabular}{|c||c|c|c|}  
\hline 
$j$ & $a_1^{(j)} ~[\mathrm{GeV}^2]$ & $a_2^{(j)} ~[\mathrm{GeV}^2]$ & $m_j~[\mathrm{GeV}]$\\ \hline
$V_1$ & $0.782$ & $-0.132$ & $1.050$\\
$V_2$ & $-4.873$ & $-0.645$ & $1.323$\\ 
$V_3$ & $3.518$ & $-0.987$ & $1.368$\\
$V_4$ & $2.243$ & $-3.813$ & $1.462$\\
$V_5$ & $-1.422$ & $3.668$ & $1.532$\\ \hline
$R_{v1}$ & $-0.985$ & $1.061$ & $2.220$\\
$R_{v2}$ & $-1.947$ & $0.551$ & $2.253$\\
$R_{v3}$ & $2.552$ & $-1.217$ & $2.256$\\
\hline
\end{tabular}
\end{center}
\end{table}
\newpage

\subsection{Acknowledgements}
We thank Young-Hui Lin and Ulf-G. Mei{\ss}ner for instructive information and correspondence about their high-precision analysis of nucleon electromagnetic form factors.  One of us (W.W.) gratefully acknowledges stimulating discussions with Jean-Paul Blaizot and Dima Kharzeev concerning the proton mass radius. 
This work has been partially supported by Deutsche Forschungsgemeinschaft (DFG grant TRR110) and National Natural Science Foundation of China (NSFC grant no. 11621131001) through the Sino-German CRC110 "Symmetries and the Emergence of Structure in QCD",  and by the DFG Excellence Cluster ORIGINS.


\end{document}